\documentclass[letterpaper,11pt ]{article}

\usepackage{graphicx}
\usepackage{bm}
\usepackage[mathlines]{lineno}

\usepackage[utf8]{inputenc}
\usepackage[T1]{fontenc}
\usepackage{mathptmx}
\usepackage{amssymb,amsmath}
\usepackage{tabularx}

\newcommand{\be}{\begin{equation}}
\newcommand{\ee}{\end{equation}}
\newcommand{\ben}{\begin{eqnarray}}
\newcommand{\een}{\end{eqnarray}}
\newcommand{\n}{\nonumber  }

\newcommand{\nd}{\noindent}

\begin{document}


\title{Virial-ans\"atze for the Schr\"odinger Equation \\ with a symmetric strictly convex potential}
\author{S. P. Flego\\
\small{ Universidad Nacional de La Plata (UNLP), Facultad de Ingenier\'{\i}a,} \\
\small{GAMEFI-UIDET, (1900) La Plata, Buenos Aires, Argentina.}}
\date{\today}
	\maketitle
				
\begin{abstract}
\nd Considering symmetric strictly convex potentials,  a local relationship  is inferred from the virial theorem, based on which a real log-concave function can be constructed. Using this as a weight function and in such a way that  the virial theorem can still be verified, 
parameter-free ans\"atze for the eigenfunctions  of the associated Schr\"odinger equation  are built.
To illustrate  the process, the technique is successfully tested against 
 the harmonic oscillator, in which it leads to the exact eigenfunctions, and against the quartic anharmonic oscillator, 
which is considered the paradigmatic testing ground for new approaches to the Schr\"odinger equation. 
\end{abstract}
\section{Introduction}
\nd As is well known, only a few quantum-mechanical models admit of exact solutions. Approximations of diverse types constitute the hard-core of the tools at the disposal of the quantum-practitioner. Some sophisticated integration approaches have been developed, as a non-exhaustive set see for instance  [1-10]. 
On the other hand, the virial theorem (VT) provides an extremely useful tool for to study a quantum-mechanical system. In its non-relativistic version, it is based on the Schr\"odinger equation (SE), and it allows making conclusions about some interesting problems without solving the SE. 
Due to this characteristic, since the 60’s, hypervirial theorems \cite{robi,castro} have been gainfully incorporated to the literature.  Also,  the subject was revisited  in the information theory context, via the strong link that exists between Fisher’s information measure and the SE [13-17].
As a direct consequence of the Legendre structure that underlies the connection between both theories, an ansatz, in terms of quadratures, 
 for the probability distribution function associated to even convex informational-potentials was derived \cite{pdf-1}, 
from which an ansatz for the ground state wave function of its associated SE was inferred. 
Then, an improved version of the procedure was derived  which can be used to deal with symmetric convex potentials \cite{pdf-2}.\\
\nd In the present communication, we treat with symmetric and strictly convex potentials in a purely quantum one-dimensional scenario. 
In this context, from the virial theorem a local relationship is inferred,  from which a real log-concave  $\chi_v$-function can be constructed. 
Then, a weight function $ \sigma_v=\chi_v^2 $ is defined  and a set of orthonormal polynomials $\{\varphi_0,\varphi_1,\cdots\}$ with respect to $ \sigma_v $, can be built. The ordered set of functions $\{\chi_0,\chi_1,\cdots\}$ where  $\chi_n=\varphi_n \chi_v$  turn out to be  ans\"atze for the eigenfunctions $\psi_n$ of the SE. The method described was designed with the goal in mind of construct a set of orthonormal  functions which satisfy the virial theorem. 
Remarkably enough, the present procedure does not involve free fitting parameters, as is common practice in other treatments. 
Clearly, once we have available ans\"atze for the eigenfunctions, approximate eigenvalues of energy  may be calculated. 
To illustrate the process, the technique is successfully tested with regards to  the harmonic oscillator, in which it leads to the exact eigenfunctions, and to the quartic anharmonic oscillator, which is considered the paradigmatic testing-ground for new approaches to the SE. 
\section[]{Preliminaries}
\subsection[]{ Convex potential}
\nd Convex potentials play an important role in many areas of science \cite{Nicu}. 
Intuitively, the graph of a convex function lies on any chord between two points on the graph. 
For a single-variable function $U(x)$, the following characterizations of convexity are helpful for the present work.
\nd If $U$ is convex on an open interval ${\cal I}$,  any local extreme of $U$ is a global minimum. Furthermore, if $U$ is strictly convex 
on ${\cal I},$  any local minimizer of $U$ is the unique strict global minimizer of $U$ on ${\cal I}.$
This is a general property of (strictly) convex function on an open set ${\cal I}$.
Then, if  $U$ is a differentiable function that achieved its minimum at $x=\xi$,  the {\it strict convexity property} can be expressed by \cite{Nicu}
	\ben \label{1}
	U(x) \text{ is strictly convex on} ~{\cal{I}}~\text{if and only if} \hspace{0.2cm}
	\forall  x \in {\cal I},~x \neq \xi~, \hspace{0.2cm} (x-\xi) ~  U\,'(x)~ > ~0. \een
where the prime denotes differentiation with respect to the function-argument. If  $U$ is at least a twice differentiable function, the property can be expressed in an equivalent form by \cite{Nicu}
\ben \label{2}
&U(x) \text{ is strictly convex on}~ {\cal{I}}~\text{if and only if} \;\hspace{0.2cm} \;U''(x)\; \geq \;0 \,&\\
&\text{and  the set of points where}~ U''(x) ~\text{vanishes does not include intervals of positive length.}&\n\een
\subsection{The quantum scenario }
Let us consider the time-independent one-dimensional  Schr\"oedinger wave equation (SE)  in dimensionless form,
 \ben \label{3} \left[-~\frac{1}{2}~\nabla_x^2~+~U(x)~\right]\psi_n(x)=E_n\psi_n(x)\:, \ \ \ \nabla_x^2~ \equiv ~\frac{d^2~}{d x^2}\een  
\nd where  $ U(x)$ stands for a real time-independent potential.

\nd In this one-dimensional scenario, the {\it virial theorem }(VT) states  that \cite{greiner}
\ben \label{4} \left\langle - ~ \nabla_x^2 ~ \right\rangle_n  =  \left\langle x \: U\,'(x) \right\rangle_n \een
 where  the expectation value is taken for stationary states of the Hamiltonian.
For one-dimensional scenarios,  $\psi_n$ is real \cite{richard},
then the VT can be written as
\ben \label{5} -\int_{-\infty}^{~\infty}{\psi_n(x) \nabla_x^2 \psi_n(x) \: dx}= ~\int _{-\infty}^{~\infty}{\psi_n^{\: 2}(x)~ x \:  U\,'(x)\:  dx} \een
\section{The virial $\chi_v$-function}
\nd For the sake of this analysis,  $U(x)$ will be assumed to be a smooth symmetric strictly convex  potential which achieved its minimum at $x=\xi$. For convenience, we can perform a shifting transformation $u=x-\xi$. Denoting with a bar the quantities referred to the new reference system $\bar{S}$,   the shifted SE is given by
\ben \label{6}\bar{ H}~\bar{\psi}_n(u)={E_n}~\bar{\psi}_n(u)~, \hspace{0.5cm} with \hspace{0.5cm}
\bar{ H}=-\: \frac{1}{2} \: \nabla_u^2~+\bar{U}(u)~,\een
 \nd where the potential and the eigenfunctions are related to the original ones by
\ben \label{7} \bar{U}(u)=U(u+\xi)=U(x) ~,\hspace{2.cm}\bar{\psi}_n(u)=\psi_n(u+\xi)=\psi_n(x) \een
\nd In $\bar{S}$, the virial theorem (\ref{5}) says that,
\ben  \label{8} -\int_{-\infty}^{~\infty}{\bar{\psi}_n(u) \nabla_u^2~\bar{\psi}_n(u)~ du}  &=& ~\int_{-\infty}^{~\infty}{ \bar{\psi}_n^{\: 2}(u)~ u \: \bar{U}\,'(u) \:~ du} \een
\nd  The left side of the above expression can be written as
\ben \label{9}
-\int_{-\infty}^{~\infty}{ \bar{\psi}_n \nabla_u^2~ \bar{\psi}_n  \ du}=
 \int_{-\infty}^{~\infty}{ \left( \bar{\psi}\,'_n\right)^2 du}=\frac{1}{4} \int_{-\infty}^{~\infty}{\bar{\psi}_n^{\:2}  \left[\left(\ln {\bar{\psi}_n^{\:2}}\right)'\right]^2  du }\een
 \nd  so from  (\ref{8}) and  (\ref{9}) one finds the following simple and convenient virial-expression
\ben  \label{10}
 \int_{-\infty}^{~\infty}{\bar{\psi}_n^2(u)\left\{\left[\left(\ln {\bar{\psi}_n^{\:2}(u)}\right)'\right]^2 - 4 \: u \: \bar{U}\,'(u) \:\right\}~ du }\:= \:0 \een

\nd By taking into account  the potential properties,  one can devise a function $\bar{\chi}$  that, by construction, verifies (\ref{10}). One merely requires fulfillment of the local relationship
\ben  \label{11}\left[\left(\ln {\bar{\chi}^{\:2}(u)}\right)'\right]^2 - 4 ~ u \: \bar{U}\,'(u) = 0\:~\een
\nd which leads to the independent solutions 
\ben \label{12} \bar{\chi}^{~\pm}(u)~ = ~exp \left(\pm \int{\sqrt{ ~ u ~\: \bar{U}\,'(u) }~ du} \right)~,  \een 
Taking into account (\ref{12}) we define the {\it virial $\bar{\chi}_v$-function}  
\ben \label{13} \bar{\chi}_{v}(u)~ = N~e^{-~\bar{g}(u)}\een
where
\ben \label{14} \bar{g}(u)=~\left\{ {\begin{array}{l}
	- \int{\sqrt{ ~ u ~\: \bar{U}\,'(u) }~ du} ~, \hspace{1.cm} if\hspace{0.2cm}~u<0 \\
+ \int{\sqrt{ ~ u ~\: \bar{U}\,'(u) }~ du} ~, \hspace{1.cm} if\hspace{0.3cm}u\geq 0\\
\end{array}} \right.  \een 
\nd and $N$ determined by the requirement that
\ben \label{15} \int_{-\infty}^{+\infty}\bar{\chi}_v^{2}(u)~du = 1~.\een

\nd  The $\chi_v$-function  (\ref{13})   provides us the tools to build useful, rather general and virially motivated ans\"atze 
$\bar{\chi}_n$ for the eigenfunction $\bar{\psi}_n$  of the SE (\ref{6}).
Before tackling this issue, we are going to show some properties of this ansatz generator.
\subsection{Properties of the  $\chi_v$-function}
\nd For sake of clarity, we are going to rewrite the properties of the potential in the $\bar{S}$-system:
\ben \label{16} 
\text{\it $\bar{U}$ is an even function:}\hspace{4.5cm}&\hspace{0.8cm}\forall u\: \in {\Re} ,\hspace{0.8cm} &\hspace{0.3cm}\bar{U}(- u)~ = ~\bar{U}(u) 
\hspace*{1.cm}\\
 \label{17}  
	\text{\it $\bar{U}$ achieved a global minimum at $u=0$:}\hspace{1.8cm}&	&\hspace{0.5cm}\bar{U}'(0)~ = ~0 \\
\label{18}
\text{\it $\bar{U}$ is a strictly convex function on}~\Re ~\text{if and only if,}\hspace{0.2cm}& 
\forall u\: \in {\Re},~u\neq 0,&\hspace{0.2cm} u ~\bar{U}'(u)~ > ~0. \\
 \label{19}
\text{\it $\bar{U}$ is a strictly convex function on}~\Re ~\text{if and only if,}\hspace{0.2cm}
& \forall u \in {\Re}\;,&\hspace{0.4cm} \;\bar{U}''(u)\; \geq \;0 \,,\hspace*{1.5cm}
\een
\hspace*{0.8cm} $\text{\it and  the set of points where} ~\bar{U}''(x) ~\text{\it vanishes does not include intervals of positive length.}$

\nd Also, due to the fact that the potential function  belongs to $\mathcal{L}_2$, it supports a serial expansion in
$\{u,\,u^2,\ldots\}$ \cite{greiner}. Taking into account that $\bar{U}$ is an even  function, the sum is over even powers, 
 \ben  \label{20}
\bar{U}(u)&=& \sum_{n\geq 1}^{}a_nu^{2n} \een
\nd  If $a_k $  is the first non-zero expansion coefficient  of $\bar{U}$, we can write
\ben  \bar{U}(u)&=& a_ku^{2k}\left[1+{\cal O}(u^2)\right] \n \een
Therefore, the asymptotically behavior of  $\bar{U}$  as $u\rightarrow 0$ is given by
\ben  \label{21}
\bar{U}(u) \sim a_ku^{2k},~\hspace{0.5cm} as \hspace{0.3cm} u \rightarrow 0 \een
with $a_k>0 $ for  the convexity property to be satisfied.


\nd We start studying the g-function defined by (\ref{14}) to then be able to approach the study of the $\chi_v$-function.

\vspace{0.3cm}

\nd{\it Main properties of  $g$-function}
\begin{itemize}
\item[$\star$]{\it $g$- domain}\\
Due to the fact that the potential is convex  (\ref{18}), $\bar{g}(u)$ is a real valued-function defined for all $u$ on $ \Re$.
\item[$\star$]{\it $g$- differentiability}\\
The potential properties allow us to conclude that the integrand that appear in the $\bar{g}$-definition (\ref{14}) is a continuous function,  then appealing to the Fundamental Theorem of Calculus and taking into account that the conditions for differentiability at $u=0$ are satisfied, the derivative of  $\bar{g}(u)$  is given by
\ben \label{22}
\bar{g}'(u) &=& \left\{ \begin{array}{l}
-{\sqrt {u\bar{U}'(u)\;}}~\hspace{0.6cm}if\hspace{0.6cm}u < 0  \\
+{\sqrt {u\bar{U}'(u)\;}}~\hspace{0.6cm}if\hspace{0.6cm}u \geq 0  \\
\end{array} \right., \een
Hence, taking into account the potential property cited,  we can conclude that $\bar{g}$ is differentiable on $\Re$.\\
The second derivative of $\bar{g}(u)$  forall $u \neq 0$ is given by
\ben \label{23} \bar{g}''(u) &=&{\frac{1}{2 |u| \sqrt{u\bar{U}'(u)}}} \left(   u \bar{U}'+ u^2 \bar{U}''\right)~,\hspace{0.5cm}\hspace{0.5cm}u \neq 0 \een
\nd Due to the fact that the potential is a strictly convex function (\ref{18}), at least twice times differentiable on $\Re$, 
we can assert that  forall  $u\neq 0$ $g(u)$ is twice times differentiable on $\Re$.  
 Hence the only point left to test is $u=0$. Considering the asymptotically behavior of  $\bar{U}$ as $ u \rightarrow 0$ (\ref{21}), we have
\ben
\int{\sqrt {u\bar{U}'(u)}du}&\sim &
 \left\{ \begin{array}{l}
-\frac{\sqrt {2k~a_k\;}}{k+1}~ |u|^{k+1}, \hspace{0.2cm} as \hspace{0.2cm} u \rightarrow 0^- \\
+\frac{\sqrt {2k~a_k\;}}{k+1}~| u|^{k+1}, \hspace{0.2cm}as \hspace{0.2cm} u \rightarrow 0^+ \\
\end{array} \right. \nonumber \een
\nd then, the behavior of the $\bar{g}$-function (\ref{14})  in the neighborhood of $u=0$ is given by
\ben \label{24} 
\bar{g}(u) &\sim&\hspace{0.2cm}\frac{\sqrt {2k a_k\;}}{k+1}~|u|^{k+1},~\hspace{0.7cm} as \hspace{0.3cm} u \rightarrow 0~\een
therefore, 
\ben  \label{25}
 \bar{g}^{''}(u) &\sim& k \sqrt {2k a_k\;}~|u|^{k-1} ,~
\hspace{0.3cm} as \hspace{0.1cm} u \rightarrow 0 \een 
 Then  forall $k \geq 1$, $\bar{g}^{''}(0^-)=\bar{g}^{''}(0^+)$.  Hence,  we can assert that $\bar{g}(u)$ is a  twice differentiable function on $\Re$.
\item[$\star$]{\it $g$- parity}\\
From (\ref{22}) , $\bar{g}'(u)$ is an odd function  in $\Re$, therefore, we can conclude that $\bar{g}(u)$ is an {\it even function},
\ben \label{26}
\bar{g}'(-u) =-\bar{g}'(u)  \hspace*{0.6cm} \longrightarrow \hspace*{0.6cm}
 \bar{g}(-u) =\bar{g}(u) ~,\hspace{0.9cm}\forall u \in \Re\hspace*{0.9cm} \een
\item[$\star$]{\it $g$- monotonicity and extreme values}\\
 Taking into account (\ref{17}) and (\ref{18}), from (\ref{22}) we have
\ben  \label{27}  
\bar{g}'(u) <0 \hspace{0.4cm} for~\hspace{0.2cm} u<0, \hspace{0.6cm} \bar{g}'(u) >0 \hspace{0.4cm} for ~\hspace{0.2cm}u >0
\hspace{0.6cm} and \hspace{0.6cm} \bar{g}'(0)=0\een
We can consequently conclude that if $u < 0$ , $\bar{g}$ is a {\it monotonically decreasing function} in the $ u$-direction, and  for $ u > 0$, $\bar{g}$  
is a {\it monotonically increasing function} in the $ u$-direction.  The unique {\it critical point} of $\bar{g}$ 
is  $u_c=0$. From the continuity and monotonicity properties of  $\bar{g}(u)$ , we can assert that it has an {\it absolute minimum value} at $u_c=0$. 
\item[$\star$]{\it $g$- curvature}\\
 From (\ref{23}) and (\ref{25}) , the second derivative of  $\bar{g}(u)$  is given by
\ben \label{28} 
\bar{g}''(u) &=& \frac{1}{2~|u|~\sqrt {u\bar{U}'(u)\;}}[u\bar{U}'(u)+u^2\bar{U}''(u)\;]~\een
Taking into account (\ref{18}) and (\ref{19}), we have
\ben \label{29}
\forall ~u \neq 0, ~u \in \Re: \hspace{0.3cm} \bar{g}''(u) >0   \een
Also, from (\ref{25}) one has $ \bar{g}''(0) \geq 0$.  Therefore we can conclude that $\bar{g}$  is a {\it strictly convex function}.
\end{itemize}
  From the above results, we can assert that:\\
\hspace*{1.cm}{\it $\bar{g}$ {\it  is a real-valued, at least twice differentiable, even and strictly convex function on } $\Re$. }


\vspace{0.5cm}

\nd {\it Main properties of  $\chi_v$-function} 

\vspace{0.3cm}

\nd Due to the fact that $\bar{g}(u)$ is a real-valued function on $\Re$,  $\bar{\chi}_v$  (\ref{13}) {\it is a positive real-valued function} on $ \Re$.
Also, taking into account that $\bar{\chi}_v$ (\ref{13}) is  the composition of  the  $\bar{g}$-function with an exponential function, we can assert that  
 $\bar{\chi}_v$ is at least twice differentiable on $\Re$. 
Furthermore, due to the fact that $\bar{\chi}_v$  is a positive real-valued function on $ \Re$ and taking into account that
$\bar{g}$ is a symmetry convex function, we can assert  that $\bar{\chi}_v$ is a symmetric log-concave function \cite{Boyd,Log} 
which satisfies the normalization condition 
$\bar{\chi}_v\stackrel{u\rightarrow \pm \infty}{\longrightarrow}0$.  We show some of inferred properties.
 \begin{itemize}
\item[$\star$]{\it $\chi_v$- parity}\\
From (\ref{13}) and (\ref{26}) we  can conclude that $\bar{\chi}_v$ {\it is an even function}.
\ben \label{30}
\bar{\chi}_v(-u) =N ~ e^{- \;\bar{g}(-u)} =N ~ e^{- \;\bar{g}(u)} =\bar{\chi}_v(u)\een
\item[$\star$]{\it $\chi_v$- monotonicity and extreme values}\\
\nd The first derivative of  $\bar{\chi}_v$ (\ref{13}) is given by 
\ben \label{31} 
 \bar{\chi}_v' (u) =~ - \bar{g}'(u)~ \bar{\chi}_v (u) \een
  Taken into account  (\ref{27}) and (\ref{31}) we conclude that
\ben \label{32}
\bar{\chi}_v'(u) > 0 ~\hspace{0.3cm} for~ u<0, \hspace{1.cm}\hspace{0.3cm}\bar{\chi}_v'(u) < 0~~\hspace{0.3cm}for~ u>0\,.\een
Therefore,   if  $u < 0$ , $\bar{\chi}_v$ {\it is a monotonically increasing function} in the $ u$-direction. 
 Also, for $ u >0$, $\bar{\chi}_v$ {\it is a monotonically decreasing function} in the $ u$-direction. 
Also,  from (\ref{27}) and (\ref{31})  the derivative of $\bar{\chi}_v$ vanishes at $u=0$, therefore the unique {\it critical point} is  $u_c=0$. 
From the continuity and monotonicity properties of $\bar{\chi}_v(u)$ (\ref{32}), we can conclude that it  has an {\it absolute maximum value} at $u=0$.
\item[$\star$]{\it $\chi_v$- curvature}\\
\nd The second derivative of  $\bar{\chi}_v$ (\ref{13}) is given by 
\ben  \label{33} 
 \bar{\chi}_v'' (u) =\left\{ \left[\bar{g}'(u)\right]^2- \bar{g}''(u) \right\}  \bar{\chi}_v (u)  \een 
\nd Considering the asymptotic behavior of $\bar{g}$  as $ u \rightarrow 0$ (\ref{24}), we have
\ben  \label{34}
 \bar{\chi}_v'' (u)  \sim \left[2k~a_k\; u^{2k}- k \sqrt{2k~a_k\;}~|u|^{k-1}\right] \bar{\chi}_v(u) 
	 \sim -  k~\sqrt{2k~a_k\;}~|u|^{k-1} \bar{\chi}_v(u) ,~\hspace{0.5cm} as \hspace{0.2cm} u \rightarrow 0\een 
	then, exist a neighborhood $(-\delta,\delta)$ of $u=0$ where $\bar{\chi}_v$ {\it is strictly concave},
\ben  \label{34a} \bar{\chi}_v'' (0) \leq 0~\hspace{0.4cm}and~
for\hspace{0.2cm}u\neq 0,~ u \in (-\delta,\delta) ~we ~have \hspace{0.4cm} \bar{\chi}_v'' (u) <0  \n  \een 

\nd Also, from (\ref{33}) and (\ref{34}) we can see that $\bar{\chi}_v'' (u)$ is a continuous function on $\Re$, and 
\nd  due to the fact that the asymptotic behavior of  $\bar{\chi}_v'' (u)$ as $ u \rightarrow \pm \infty$ is dominated by $\bar{\chi}_v (u)$ 
  we can assert that the second derivative of the $\bar{\chi}_v$-function converges to zero at $\pm \infty$. 
Then,  Rolle's theorem says that the second derivative vanishes at least two times on $\Re$. 
These inflexion points $u_r$ can be determinate from (\ref{33}) imposing $\bar{\chi}''(u_r)=0$, that lead to require that 
\ben  \label{35}
\left. \left[\bar{g}'(u)\right]^2 - \bar{g}''(u)~\right|_{u=u_r}=0   \een  
Due to the fact that  $ \bar{\chi}_v'' (u)$ is an even function, namely, 
 \ben  \label{36} 
 \bar{\chi}_v'' (-u)=\left\{[\bar{g}'(-u)]^2-  \bar{g}''(-u)\right\}\bar{\chi}_v (-u)=
 \left\{ [\bar{g} '(u)]^2-\bar{g} ''(u)\right\}\bar{\chi}_v (u) = \bar{\chi}_v'' (u) \een
 and taking into account that $\bar{\chi}_v$-function has a maximum in the coordinate origin, the inflection points occur in pairs and they are located on each side of the center of symmetry $u=0$  and at the same distance from it.
It is noteworthy that we expect that there will only be one pair of them, since, as we will see, such points are closely related to the classical return points of the theory. If this condition is met, the graph of $\bar{\chi}$ looks like  a bell-shaped curve.
\end{itemize}
\nd From the  above analytical study we can conclude that: 
\begin{center}
 {\it $\bar{\chi}_v$ is a log-concave real-valued function, at least twice differentiable on $\Re$. \\
Also, it is an even  function which converges to zero at $\pm \infty$.}
\end{center}
\section{Ans\"atze  for the eigenstates of the Schr\"odinger Equation}
\subsection{The  ans\"atze for the eigenfunctions}
\nd With the goal in mind of retaining the original properties of the potential, we immediately note that excellent results could eventually be achieved if we used
 the virial  $\bar{\chi}_v$-function (\ref{13}) to build  ans\"atze  $\bar{\chi} _n$ for the eigenfunctions $\bar{\psi} _n$ of the SE (\ref{6}). With this purpose, we define the
{\it virial weight function} as
\ben \label{37}
 \bar{\sigma}_v(u) \equiv \bar{\chi_v}^2(u)=N^2~e^{-2 \bar{g}(u)}\een
which obviously has the same properties of the $\bar{\chi_v}$-function and  therefore is a good candidate to be an ansatz for the probabilities density function.
Then, we chose an {\it appropriate set} $\{\bar{\varphi}_n, n=0,1,2...\}$ of orthonormal functions  with weight function $\bar{\sigma}_v(u)$ \cite{szego},
\ben \label{38}
 \left( {\bar{\varphi}_i ,\bar{\varphi}_j } \right)_{\sigma_v}  =  \int_{-\infty}^{~\infty}{\bar{\varphi}_i(u)\bar{\varphi}_j(u)\bar{\sigma}_v(u)\;du}=\delta_{ij},
 \hspace{1.5cm} \bar{\sigma}_v(u) \equiv \bar{\chi}_v^2(u)\een
\nd and propose the normalized ansatz $\bar{\chi} _n$ for the eigenfunction $\bar{\psi} _n$    as
\ben \label{39}
\bar{\chi}_n(u)=\bar{\varphi}_n(u) \bar{\chi}_v(u)~,\hspace{2.cm}n=0,1,2,...\een
\ben \label{40}
\left\langle {\bar{\chi}_i | \bar{\chi}_j } \right\rangle  =\int_{-\infty}^{~\infty}{\bar{\chi}_i (u)\bar{\chi}_j(u) \;du} =
 \int_{-\infty}^{~\infty}{\bar{\varphi}_i(u)\bar{\varphi}_j(u)\bar{\chi}_v^2(u)\;du}=
\left( {\bar{\varphi}_i ,\bar{\varphi}_j } \right)_{\sigma_v}= \delta_{ij} \een
\nd To obtain the sequence $\{ \bar{\varphi} _n\} $ we need to choose a basis $\left\{\bar{\nu}_0,\bar{\nu}_1,\bar{\nu}_2\,...\right\}$.
 {\it For determine which is the {\it appropriate set}  $\{\bar{\nu}_k, k=0,1,2,\cdots\}$ one looks the {\it Virial theorem}.}
\begin{itemize}
\item[ ]  We started from the expectation value of the kinetic term
 \ben \label{41}
 \noindent \left\langle  \bar{\chi}_n \left|\nabla_u^2~ \right|   \bar{\chi}_n \right\rangle
 &=&\int_{-\infty}^{~\infty}{ \bar{\chi}_n \nabla_u^2~\bar{\chi}_n} ~du=~-\int_{-\infty}^{~\infty}{ \left(  \bar{\chi}\,'_n\right)^2du} =\n\\
&=&-\int_{-\infty}^{~\infty}{\left(  \bar{\varphi}\,'_n~\bar{\chi}_v+\bar{\varphi}_n~  \bar{\chi}\,'_v \right)^2du} =\n\\
&=&-\int_{-\infty}^{~\infty}{\left[ \left(  \bar{\varphi}\,'_n \right)^2\bar{\chi}_v^2+
2\bar{\varphi}_n  \bar{\varphi}\,'_n \: \bar{\chi}_v \bar{\chi}\,'_v +\bar{\varphi}_n^2 \! \left(  \bar{\chi}\,'_v\right)^2 \right]du } \een 
\nd Integration by parts of the second  term on the right side of the above expression lead to
\ben \label{42}
\int_{-\infty}^{~\infty}{2\bar{\varphi}_n  \bar{\varphi}\,'_n ~  \bar{\chi}_v \bar{\chi}\,'_v~du}= 
~- \int_{-\infty}^{~\infty}{ \left(\bar{\varphi}_n  \bar{\varphi}\,'_n\right)' ~  \bar{\chi}_v^2~du}  \hspace{1.5cm}\een 

\nd Substituting  (\ref{42}) in (\ref{41}), we obtain
\ben \label{43}
\noindent \left\langle  \bar{\chi}_n \left|\nabla_u^2~ \right|   \bar{\chi}_n \right\rangle
&=&-\int_{-\infty}^{~\infty}{ \left\{ \left[ \left( \bar{\varphi}\,'_n\right)^2
-\left(\bar{\varphi}_n  \bar{\varphi}\,'_n\right)'\,\right]  \bar{\chi}_v^2 +\bar{\varphi}_n^2 \! \left(  \bar{\chi}\,'_v\right)^2 \right\}du}\n\\
   &=&\int_{-\infty}^{~\infty}{ \left[ \left( \bar{\varphi}_n  \bar{\varphi}\,''_n\right) \:  \bar{\chi}_v^2 -\bar{\varphi}_n^2 \! \left(  \bar{\chi}\,'_v\right)^2 \right]du}  \een 
Taking into account (\ref{22}), (\ref{31}) and (\ref{39}) we can write
\ben \label{44}
\bar{\varphi}_n^2 \! \left( \bar{\chi}\,'_v\right)^2 =[g'(u)]^2\: \bar{\varphi}_n^2 \bar{\chi}_v^2=u \: \bar{U}\,'(u)  \:\bar{\chi}_n^2
 \een 

\nd Substituting (\ref{44}) in (\ref{43}), we obtain
\ben \label{45a}
\noindent \left\langle  \bar{\chi}_n \left|\nabla_u^2~ \right|   \bar{\chi}_n \right\rangle
=\int_{-\infty}^{~\infty}{ \left[ \left( \bar{\varphi}_n  \bar{\varphi}\,''_n\right) \:  \bar{\chi}_v^2 -u \: \bar{U}\,'(u)  \:\bar{\chi}_n^2 \right]du}\n  \een 
which can be written as
\ben \label{45}
\noindent \left\langle  \bar{\chi}_n \left|\nabla_u^2~ \right|   \bar{\chi}_n \right\rangle
 = \left(  \bar{\varphi}_n ,\bar{\varphi}\,''_n \right)_{\sigma_v} -
 \noindent \left\langle  \bar{\chi}_n \left|u \: \bar{U}\,'(u) \right|   \bar{\chi}_n \right\rangle
\een 
\nd We immediately note that $\bar{\chi}_n=\bar{\varphi}_n~\bar{\chi}_v$ satisfies the virial theorem (\ref{8}) if $\bar{\varphi}_n$  satisfies the following virial condition
\ben \label{46}
\left(  \bar{\varphi}_n ,\bar{\varphi}\,''_n \right)_{\sigma_v} =~0
 \een 
\nd The above result suggests that  $\{\bar{\varphi}_n\; , n = 0, 1, 2, ...\}$ can be choose as a  family of  orthonormal real polynomials associated with the weight function $\bar{\sigma}_v(u)$,
\ben \label{47} 
\bar{\varphi}_n(u)&=&  \sum_{j=0}^{n}{\alpha_{nj}~u^j}\een
with the coefficients $\alpha_{nj}$ determined by the condition (\ref{38}).
\nd The {\it virial condition} (\ref{46}) is satisfied due to the fact that the second derivative of $\bar{\phi}_n$ can be written as a linear combination 
of polynomial $\bar{{\phi}}_k$ of degree equal or less than $(n-2)$, then  it is orthogonal to the polynomial $\bar{{\phi}}_{n}$. 
Therefore, an appropriate basis is $\{\bar{\nu}_k=u^k, k=0,1,2,\cdots\}$.
\end{itemize}

\vspace{0.3cm}

\nd The polynomial sequence $\{\bar{{\varphi}}_n,\; n=0,1,2,...\}$ can be obtained using the Gram-Schmidt orthonormalization  process \cite{szego,chih},
\ben \label{48}
\bar{\varphi}_o(u)&=&1\nonumber\\
\bar{\varphi}_n(u)&=&a_n\left[u^n-\sum_{k=0}^{n-1}{\left(  u^n ,  \bar{\varphi}_k \right)_{\sigma_v}}~\bar{\varphi}_k(u)\right] , \hspace{0.4cm} n\geq1
\een
where   $a_n $ are constants determined by the normalization condition. They can be expressed in terms of the Gram determinant \cite{szego}. Note that, from (\ref{39}) and (\ref{48}) we have 
\ben  \label{49} \bar{\chi}_o=\bar{\chi}_v  ~, \hspace{1.5cm}\bar{\chi}_n(u)=\bar{\varphi}_n(u) \bar{\chi}_o(u)~\een
then, we can write
\ben \label{50}
 \left( {\bar{\varphi}_i ,\bar{\varphi}_j } \right)_{\sigma_v}  =  
\left\langle \bar{\chi}_o \left|{\bar{\varphi}_i \bar{\varphi}_j }\right|   \bar{\chi}_o \right\rangle \equiv
\left\langle {\bar{\varphi}_i \bar{\varphi}_j } \right\rangle_{_0}~,  \hspace{1.cm}
\left( {u^n, \bar{\varphi}_k}  \right)_{\sigma_v}  = 
\left\langle  \bar{\chi}_o \left| {u^n~\bar{\varphi}_k}\right|   \bar{\chi}_o \right\rangle \equiv
\left\langle {u^n \bar{\varphi}_k} \right\rangle_{_0}  \een

\vspace{0.5cm}

\noindent $\star~${\it The virial eigenfunctions in the $S$-system}.

\nd Finally, the inverse shifting transformation $x=u+\xi$ leads to the original referential system $S$. From (\ref{13}),  (\ref{14}),  (\ref{39})  and (\ref{47}), the desired ans\"atze for the eigenfunctions are given by
\ben \label{51}
\noindent \chi_n(x)=\varphi_n(x) \chi_v(x)~, \een
where
\ben \label{52}
 \chi_v(x)=N~ e^{-g(x)} ~, \hspace{1.4cm}\varphi_n(x) =  \sum_{j=0}^{n} {\alpha_{nj}~(x-\xi)^j }~,\een
with 
\ben \label{53}  {g}(x)= \left\{ {\begin{array}{l}
	- \int{\sqrt{(x-\xi)\, U\,'(x)} dx}  \hspace{0.8cm} if\hspace{0.2cm}~x<\xi\\
  + \int{\sqrt{(x-\xi)\,   U\,'(x)} dx}  \hspace{0.8cm} if\hspace{0.3cm}x\geq \xi\\
\end{array}} \right.  \een 
and the coefficients $\alpha_{nj}$ determined by the condition (\ref{38}).
\subsection{The approximate energy  eigenvalues}
\nd Once we have at our disposal the ans\"atze for the eigenfunctions,  we can obtain approximate energy eigenvalues,
\ben  \label{54}
E_n  & \approx & E_n^{ans}= \left\langle \chi_{n} \left| H \right|  \chi_{n} \right\rangle \een
Also, using the virial theorem, we can write
\ben  \label{55}
 E_n^{ans}=  \left\langle \chi_{n} \left| \frac{1}{2} ~(x-\xi)~U'(x)+U(x) \right| \chi_{n} \right\rangle \een
\section{Applications}
\nd  As  illustrations of  the procedure, we deal below with  the harmonic oscillator and with the quartic anharmonic oscillator.
 \subsection{The Harmonic Oscillator (HO) } 
\nd The Schr\"odinger equation for a particle of unit mass in a shifted
harmonic potential reads,
\be \label{HO-1}
\left[-~\frac{1}{2}~\nabla_x^2~+~\frac{1}{2}~\omega^2~ (x-\xi)^2~\right]\psi_n~= ~ E_n~\psi_n~, \ee
{\it We start focusing on the potential and its derivatives,}
\ben  \label{HO-2}
U(x)= \frac{1}{2} \omega^2 (x-\xi)^2,\hspace{1.cm}U'(x)=  \omega^2 (x-\xi) \;,\hspace{1.cm}U''(x)= \omega^2\;. \een
\nd Immediately we observe that the potential $U$ is a symmetric function with an unique minimum  at $x=\xi$,
\ben  \label{HO-3}
\forall x \in \Re~, \hspace{0.5cm} U(2\xi-x) = U(x)~, \hspace{1.cm}  U'(\xi)=0  ~,\hspace{0.5cm}   U''(\xi)=\omega^2 > 0 \een
Also, $U$ is a strictly convex function (\ref{1}),
\ben  \label{HO-4}
\forall x\neq\xi, ~x \in \Re~, \hspace{1.cm}   (x-\xi)~U'(x)=  ~\omega^2 (x-\xi)^2 > 0 \een
\nd The characteristics of the potential allow applying the technique presented in the previous section. The virial $\chi_v$-function [see (\ref{52}) and (\ref{53})] takes the form
\ben \label{HO-5}  {\chi}_{v}(x)=
\left\{ {\begin{array}{l}
	N\, exp \left(+ \int{\sqrt{\omega^2 (x-\xi)^{2}} dx} \right) \hspace{0.8cm} if\hspace{0.2cm}~x<\xi\\
  N\, exp \left(- \int{\sqrt{\omega^2 (x-\xi)^{2}} dx} \right) \hspace{0.8cm} if\hspace{0.3cm}x\geq \xi\\  \end{array}} \right.  \een 
An elementary integration leads to
\ben \label{HO-6}
{\chi}_{v}(x)~=~N~exp \left[-~\frac{\omega}{2}(x-\xi)^{2}\right].\een
\nd Enforcing the normalization condition, we obtain
\ben \label{HO-7}
\chi_{v}(x)~=~\left(\frac{\omega}{\pi}\right)^{1/4}~exp \left[-\frac{\omega}{2} (x-\xi)^{2}\right], \een 
which coincides with the exact ground state eigenfunction of the HO. The virial weight function is given by
\ben \label{HO-8}
\sigma_v(x)~=~\chi_{v}^2(x)~=~\left(\frac{\omega}{\pi}\right)^{1/2}~exp\left[-{\omega} (x-\xi)^{2}\right]. \een
\nd For  calculate the ans\"atze $\{\chi_n\}$ for the eigenfunctions $\{\psi_n\}$ we need a  real family of orthonormal polynomials $\{ \varphi_n\}$ associated with the weight function $\sigma_v (x)$ (\ref{HO-8}),
\ben \label{HO-9}
\left(   {\varphi}_n, {\varphi}_m\right)_{\sigma_v} 
=\left(\frac{\omega}{\pi}\right)^{1/2}\int_{-\infty}^{+\infty}{{\varphi}_n(x) {\varphi}_m(x) ~exp\left[-{\omega} (x-\xi)^{2}\right]~dx}=\delta_{nm}\een
A close look at the orthonormalization condition allows us to relate $\varphi_n$ to the Hermite polynomials $H_n$ \cite{szego}. 
Making the variable change $ v = \sqrt{\omega} (x-\xi) $ in (\ref{HO-9}) we have
\ben \label{HO-10}
\left(\frac{\omega}{\pi} \right)^{1/2}\int_{-\infty}^{+\infty}{{\varphi}_n(v/ \sqrt{\omega}+\xi){\varphi}_m(v/ \sqrt{\omega}+\xi) ~exp\left[-v^2\right]~\frac{dv}{\sqrt{\omega}}}
&=& \delta_{nm}\een
Then, we can identify  \cite{szego}  
\ben \label{HO-11}
{\varphi}_n(v/\sqrt{\omega}+\xi)=\frac{H_n(v)}{\sqrt{2^n~n!~}}\hspace{0.3cm}\longrightarrow \hspace{0.3cm}
{\varphi}_n(x)=\frac{1}{\sqrt{2^n~n!~}}H_n(\sqrt{\omega}(x-\xi))\een
Finally, from (\ref{51}),  (\ref{HO-7}) and (\ref{HO-11}), the  ans\"atze $\chi _n$ for the eigenfunctions $\psi _n$   are given by  
\ben \label{HO-12}
\chi_{n}(x)~=~\left(\frac{\omega}{\pi}\right)^{1/4}~\sqrt{\frac{1}{2^n ~n!}}~H\left( \sqrt{\omega}~(x-\xi)\right)
~exp\left[-\frac{\omega}{2} (x-\xi)^{2}\right], \een
In this form, as we can see, the virial treatment leads  to the exact eigenfunctions of the harmonic oscillator.
\subsection{The Quartic Anharmonic Oscillator (AHO)}
\nd The Schr\"odinger equation for a particle of unit mass in a shifted
quartic anharmonic potential reads
 \be \label{AHO1}
\left[-~\frac{1}{2}~\nabla_x^2~
+~\frac{1}{2}~\omega^2~ (x-\xi)^2~+ ~\lambda ~(x-\xi)^4 \right]\psi_n~=
~ E_n~ \psi_n,  \hspace{1.cm} \lambda \geq 0\ee 
\nd where $\lambda$ is the anharmonicity constant.

\vspace{0.5cm}

\nd {We start focusing on the potential properties}. 
\nd The derivatives of the potential are given by
\ben  \label{AHO2}
U'(x)= \omega^2 (x-\xi)~+ 4\lambda ~(x-\xi)^3 , \hspace{0.5cm}U''(x)= \omega^2+ 12\lambda ~(x-\xi)^2 \een
\nd Intermediately we observe that the potential $U$ is  a symmetric function with an unique minimum  at $x=\xi$,
\ben  \label{AHO3}
\forall x \in \Re~, \hspace{0.4cm} U(2\xi-x) = U(x)~, \hspace{0.5cm}  U'(\xi)=0  ~,\hspace{0.5cm}   U''(x) > 0 \een
Also, $U$ is a strictly convex function  (\ref{1}), 
\ben  \label{AHO4}
\forall x\neq\xi, ~x \in \Re~, \hspace{0.5cm}   (x-\xi)~U'(x)= \omega^2 (x-\xi)^2+ 4 \lambda (x-\xi)^4  > 0 \een
\nd Then, the characteristics of the potential allow applying the technique cited in the previous section. We proceed to do it.

\vspace{0.4cm}

\nd{ \large $\star$ \it The virial $\chi_v$-function.}

\vspace{0.4cm}

\nd Substituting (\ref{AHO4}) in (\ref{53}), realizing  an elemental integration and  enforcing that ${g}$ remain finite in the limit in which 
$\lambda \rightarrow 0$, we obtain
\ben \label{AHO5}
g(x)~=-~\frac{\omega^3}{12\lambda}\left[1-
 \left(1+ \frac{4 \lambda}{\omega^2} ~ (x-\xi)^{2}\right)^{3/2}\right]~,
\een 
and the $\chi_v$-function (\ref{52}) is given by
\ben \label{AHO6}
\chi_v (x) =N~exp\left[-{g(x)}\right]=N~exp\left\{\frac{\omega^3}{12\lambda}\left[1-
 \left(1+ \frac{4 \lambda}{\omega^2}  (x-\xi)^{2}\right)^{3/2}\right]\right\},\een
\nd where $N$ is determined by the normalization condition. 
\vspace{0.4cm}

\nd{ \large $\star$ \it The ans\"atze for the eigenfunctions}

\vspace{0.2cm}
\nd For  calculate the ans\"atze $\{\chi_n\}$  we need a  real family of orthonormal polynomials $\{ \varphi_n\}$ associated with the weight function 
$\sigma_v (x) $ given by
\ben \label{AHO7}
\sigma_v (x) =\chi_{v}^2(x)=N^2~exp\left\{\frac{\omega^3}{6\lambda}\left[1-
 \left(1+ \frac{4 \lambda}{\omega^2}  (x-\xi)^{2}\right)^{3/2}\right]\right\},\een
\nd Explicit expressions for the orthonormal polynomials $\{ \varphi_n\}$, in terms of  the moments $\langle (x-\xi)^n\rangle_{_0}$ can be obtained using the Gram-Schmidt process (\ref{48}).  
\nd Also, it is computationally advantageous to express ${{\varphi}}_n$ in terms of lower- order orthogonal polynomials
 using the Christoffel-Darboux recurrence formula \cite{szego}. Then we have
\ben \label{TTR}
{\varphi}_{0}(x)&=&1 \n  \\
{\varphi}_{1}(x)&=& \left\langle{(x-\xi)^2}\right\rangle_{_0}^{- 1/2}~(x-\xi)  \\
{\varphi}_{n}(x)&=&\beta_{n}\left[(x-\xi)~{\varphi}_{n-1}(x)-\left\langle{(x-\xi)~{\varphi}_{n-1}{\varphi}_{n-2}}\right\rangle_{_0}~{\varphi}_{n-2}(x)\right] ~,\hspace{0.7cm}n\geq2\n
\een
where
\ben \label{beta-n}
\beta_{n}= \left[ \left\langle{(x-\xi)^2{\varphi}_{n-1}^{~2}}\right\rangle_{_0}
-\left\langle{(x-\xi){\varphi}_{n-1}{\varphi}_{n-2}}\right\rangle_{_0}^2 \right]^{-\frac{1}{2}}
\een
\ben
\left\langle{(x-\xi)^k{\varphi}_{i}{\varphi}_{j}}\right\rangle_{_0}= \int_{-\infty}^{~\infty}{(x-\xi)^k{\varphi}_{i}(x){\varphi}_{j}(x)~\sigma_v(x)~dx}
\een
\nd Finally, the ans\"atze $\chi_n$ for the eigenfunctions $\psi_n$ of the SE (\ref{AHO1}) are given by [see (\ref{51})]
\ben \label{AHO8}
\chi_n (x) =\varphi_n(x)~\chi_v (x) =N~\varphi_n(x)~exp\left\{\frac{\omega^3}{12\lambda}\left[1-
 \left(1+ \frac{4 \lambda}{\omega^2}  (x-\xi)^{2}\right)^{3/2}\right]\right\}\een


\vspace{0.4cm}

\nd{ \large $\star$  \it The approximate energy eigenvalues}

\vspace{0.4cm}
\nd  The approximate eigenenergies  (\ref{55}) are given by
\ben \label{AHO9}
 E^{ans}_n = \left\langle \chi_n \left| \omega^2(x-\xi)^2+3 \lambda  (x-\xi)^4\right|  \chi_n \right\rangle 
\een
where the expectation values are taken respect to the ansatz $\chi_n$ given by (\ref{AHO8}).

\vspace{0.6cm}

\nd{\large $\star$  \it Results and Discussion}

\vspace{0.4cm}
\nd The reader may pass judgement below on the accuracy of  the procedure.
 The  curves in  Figure 1 corresponding to a shift $\xi=4$, $\omega=1$,  and several values of the anharmonicity-constant $\lambda$.
 In each graph  we plotted both the first five eigenfunctions, for a given $\lambda$-value, obtained via a numerical approach to the SE (Matslise program was used) and the corresponding ans\"atze obtained using (\ref{AHO8}). The graphs corresponding to the numerical solutions are plotted with dashed black lines and the ans\"atze are plotted with solid lines
		(black for $\chi_o$, red for $\chi_1$, blue for 		$\chi_2$, green for $\chi_3$ and coral for $\chi_4$). 
 As can be seen, for small $\lambda$-values the curves  overlap 
but, as $\lambda$ increases, the curve associated with the ground state ansatz-eigenfunction grows a little wider
  and falls more gradually than the curve corresponding to the exact one. This difference propagates the excited states. 

\vspace{0.3cm}

\nd Table 1 shows the energy eigenvalues (\ref{AHO9}) corresponding to the first five eigenstates of the AHO, corresponding to the eigenfunctions shown in figure 1.  In each table, the values of the first column correspond to the principal quantum number.  The values of the second column correspond to the energy eigenvalues obtained  via a numerical approach to the SE (Matslise program was used) and correspond to those  that one finds in the literature.  The values, in the third column correspond to the approximate energy eigenvalues obtained instead via the present parameter-free procedure.
The fourth and fifth columns display respectively  the associated  error $\Delta_n$, and the percent relative error $\epsilon_n$, 
\ben \label{error}
\Delta_n={E^{ans}_n-E^{}_n} ~, \hspace{1.cm}
\epsilon_n=\frac{E^{ans}_n-E^{}_n}{ E^{}_n}~100
\een

\nd	As can be seen, for each eigenstate with principal number $n$, the percent relative error $\epsilon_n$ depends on $\lambda$-values. 
For small  $\lambda$-values, $\epsilon_n$ is small and tends to zero when $\lambda$ tends to zero, 
which is consistent with the fact that the potential tends to the harmonic oscillator for which the procedure leads to the exact  solutions. 
When $\lambda$ increases, $\epsilon_n$ increases.
Nevertheless, for all  values of $\lambda$ considered, the ansatz-results are in good agreement with  the numerical ones.
 Of course, whether this is sufficiently accurate depends on the requirements.
	\newpage
\begin{center}
{\bf Figure 1. The first five eigenfunctions  and the corresponding ans\"atze \\
\hspace*{0.9cm} for the quartic AHO for several parameter values. }
 
\vspace{0.3cm}

\includegraphics[width=16.0cm,height=21.cm,angle=0]{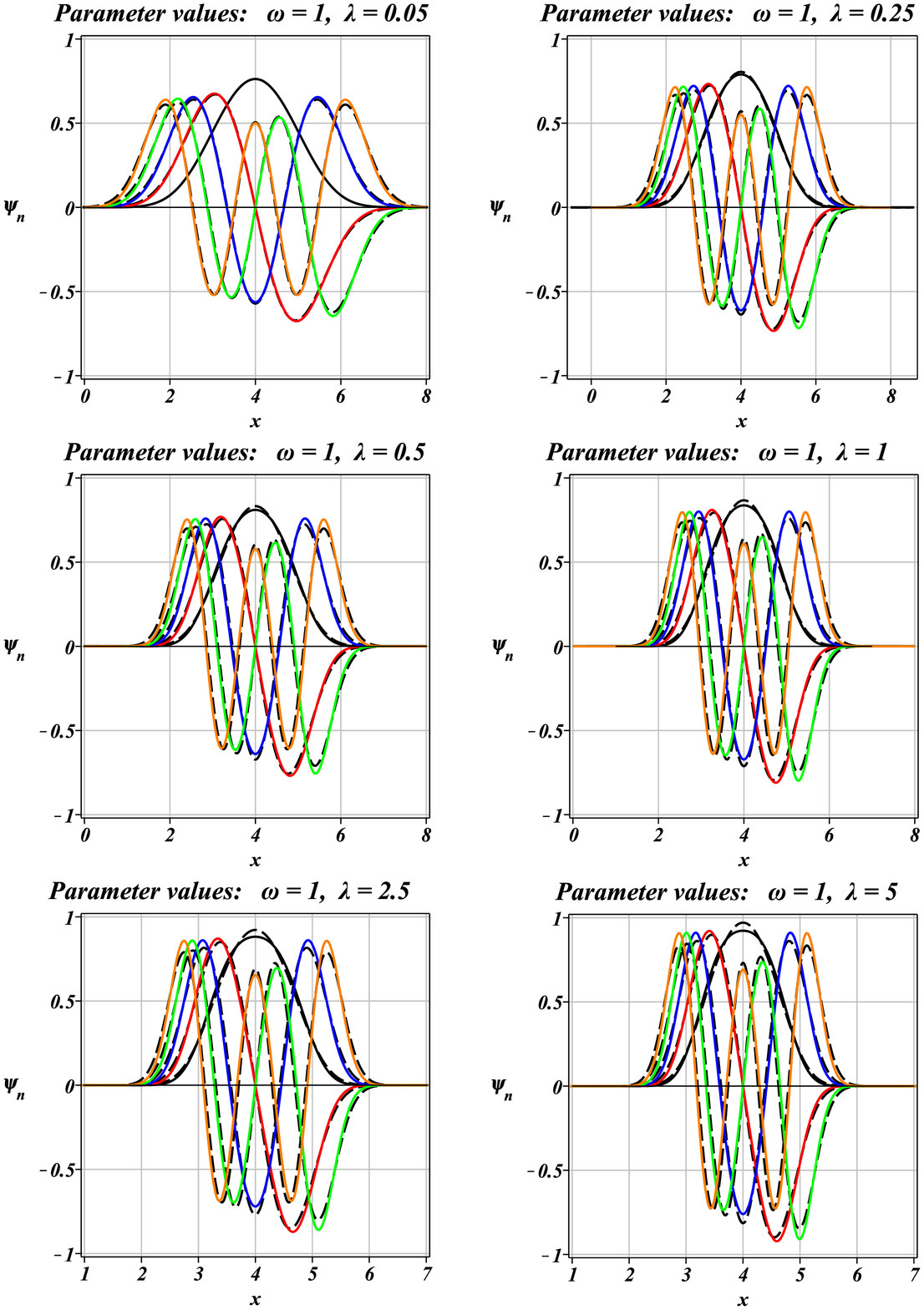}
\end{center}

\newpage
\begin{center}
\nd {\bf  Table 1. Energy eigenvalues for   the AHO for several  $\lambda$-values \hspace*{0.7cm}}\\

\vspace{0.6cm}
\footnotesize
\begin{minipage}[t]{ 8.0cm}
\begin{tabular}{|c|r|r|r|r|}
\hline
\multicolumn{5}{|c|}{\bf  Parameter values: $\omega=1$,  $\lambda=0.05$}\\
\hline
$\, n \, $&$E_n^{}\hspace{0.2cm}$&$E_n^{ans}\hspace{0.2cm}$&$\Delta_n\hspace{0.2cm}$&$~\epsilon_n ~\%    $\\
\hline
\hline
0&0.53264275&0.53305374&0.00041100&0.08\\
1&1.65343600&1.65504451&0.00160850&0.10\\
2&2.87397965&2.87793942&0.00395977&0.14\\
3&4.17633892&4.18414692&0.00780801&0.19\\
4&5.54929781&5.56234990&0.01305209&0.24\\
5& ~  6.98496312& ~  7.00456495&0.01960183&0.28\\
\hline
\end{tabular}

\end{minipage}
\begin{minipage}[t]{ 8.0cm}
\begin{tabular}{|c|r|r|r|r|}
\hline
\multicolumn{5}{|c|}{\bf  Parameter values: $\omega=1$,  $\lambda=0.25$}\\
\hline
$\, n \, $&$E_n^{}\hspace{0.2cm}$&$E_n^{ans}\hspace{0.2cm}$&$\Delta_n\hspace{0.2cm}$&$~\epsilon_n ~\%    $\\
\hline
\hline
0&0.62092703&0.62390385&0.00297682&0.48\\
1&2.02596616&2.03517802&0.00921186&0.46\\
2& 3.69845032&3.71846350&0.02001318&0.54\\
3&5.55757714&5.59366961&0.03609247&0.65\\
4&7.56842288&7.62347161&0.05504873&0.73\\
5&~   9.70914789& ~   9.78622249&0.07707460&0.79\\
\hline
\end{tabular}
\end{minipage}

\vspace{0.5cm}
\begin{minipage}[t]{ 8.0cm}
\begin{tabular}{|c|r|r|r|r|}
\hline
\multicolumn{5}{|c|}{\bf  Parameter values: $\omega=1$,  $\lambda=0.5$}\\
\hline
$\, n \, $&$E_n^{}\hspace{0.2cm}$&$E_n^{ans}\hspace{0.2cm}$&$\Delta_n\hspace{0.2cm}$&$~\epsilon_n ~\%    $\\
\hline
\hline
0&0.69617582&0.70188134&0.00570552&0.82\\
1&2.32440635&2.34037539&0.01596904&0.69\\
2&4.32752497&4.36091392&0.03338894&0.77\\
3&6.57840195&6.63697255&0.05857061&0.89\\
4&9.02877872&9.11533908&0.08656035&0.96\\
5&  11.64872074&  11.76760219&0.11888145&1.02\\
\hline
\end{tabular}
\end{minipage}
\begin{minipage}[t]{ 8.0cm}
\begin{tabular}{|c|r|r|r|r|}
\hline
\multicolumn{5}{|c|}{\bf  Parameter values: $\omega=1$,  $\lambda=1$}\\
\hline
$\, n  $&$E_n^{}\hspace{0.2cm}$&$E_n^{ans}\hspace{0.2cm}$&$\Delta_n\hspace{0.2cm}$&$~\epsilon_n ~\%    $\\
\hline
\hline
0&0.80377065&0.81363891&0.00986826&1.23\\
1&2.73789227&2.76315528&0.02526302&0.92\\
2&5.17929169&5.23069405&0.05140237&0.99\\
3&7.94240398&8.03065053&0.08824655&1.11\\
4&10.96358310&11.09072670&0.12714360&1.16\\
5&  14.20313912&  14.37547180&0.17233268&1.21\\
\hline
\end{tabular}
\end{minipage}

\vspace{0.5cm}
\begin{minipage}[t]{ 8.0cm}
\begin{tabular}{|c|r|r|r|r|}
\hline
\multicolumn{5}{|c|}{\bf  Parameter values: $\omega=1$,  $\lambda=2.5$}\\\hline
$\, n  $&$E_n^{}\hspace{0.2cm}$&$E_n^{ans}\hspace{0.2cm}$&$\Delta_n\hspace{0.2cm}$&$~\epsilon_n ~\%    $\\
\hline
\hline
0&1.00917032&1.02710883&0.01793851&1.78\\
1&3.50673959&3.54850320&0.04176361&1.19\\
2&6.73386520&6.81699876&0.08313356&1.24\\
3&10.40698348&10.54673126&0.13974778&1.34\\
4&14.43749818&14.63396937&0.19647119&1.36\\
5&  18.76940764&  19.03290198&0.26349434&1.40\\
\hline
\end{tabular}
\end{minipage}
\begin{minipage}[t]{ 8.0cm}
\begin{tabular}{|c|r|r|r|r|}
\hline
\multicolumn{5}{|c|}{\bf  Parameter values: $\omega=1$,  $\lambda=5$}\\
\hline
$\, n  $&$E_n^{}\hspace{0.2cm}$&$E_n^{ans}\hspace{0.2cm}$&$\Delta_n\hspace{0.2cm}$&$~\epsilon_n ~\%    $\\
\hline
\hline
0&1.22458703&1.25080186&0.02621482&2.14\\
1&4.29950172&4.35732149&0.05781977&1.35\\
2&8.31796074&8.43202280&0.11406206&1.37\\
3&12.90313811&13.09266465&0.18952654&1.47\\
4&17.94258562&18.20570296&0.26311734&1.47\\
5&  23.36454046&  23.71576456&0.35122410&1.50\\
\hline
\end{tabular}
\end{minipage}
\end{center}
\normalsize

\vspace{0.3cm}

\nd From the above results, one immediately note that  the free-parameter ans\"atze, which satisfy the virial theorem and incorporate the  symmetries of the potential, exhibit the behavior of the system with its intrinsic parameters. Nevertheless, for practical applications one needs to know an estimate of the degree of approximation of the results. Studies in this line  are in progress and they will be reported elsewhere. 

\section{ Conclusions}
\nd In this communication we  dealt with real, one-dimensional, symmetrical and strictly convex potentials, quite a large family indeed.
 For these potentials, the virial theorem was employed so as to infer, via symmetry considerations, an ordered set of orthonormal  functions which turn out to be ans\"atze for the eigenfunctions of the stationary Schr\"odinger  equation corresponding to the potential in question. The procedure is elementary. Further comments about the method itself are unnecessary. Despite its simplicity and parameter-free character, it provides good results, as evidenced by  the examples  here examined. Thus, the well-known exact solutions of the harmonic oscillator can be reproduced, and a very good approximation to the numerical solutions of the quartic anharmonic oscillator, an old problem of quantum mechanics, can  be determined. 

\nd To sum up, in this communication  a new general parameter-free procedure was introduced in order to obtain ans\"atze for eigenvalue-problems of linear operators with symmetric strictly convex potentials,  whose use seems  to constitute a promising approach, given the results presented here.

\vspace{1.cm}

\nd {\bf Acknowledgment-} Partial support from the program "Investigaci\'on en el \'area de Ciencias B\'asicas, Facultad de Ingenier\'{\i}a" (UNLP, Argentine) is acknowledged.


\end{document}